# Ultrafast transition between exciton phases in van der Waals heterostructures

P. Merkl[1], F. Mooshammer[1], P. Steinleitner[1], A. Girnghuber[1], K.-Q. Lin[1], P. Nagler[1], J. Holler[1], C. Schüller[1], J. M. Lupton[1], T. Korn[1,3], S. Ovesen[2], S. Brem[2], E. Malic[2*] and R. Huber[1*]

[1]Department of Physics, University of Regensburg, D-93040 Regensburg, Germany.
[2]Department of Physics, Chalmers University of Technology, SE-41258 Gothenburg, Sweden.
[3]Current address: Department of Physics, University of Rostock, D-18059 Rostock, Germany.

**Heterostructures of atomically thin van der Waals bonded monolayers have opened a unique platform to engineer Coulomb correlations, shaping excitonic[1-3], Mott insulating[4], or superconducting phases[5,6]. In transition metal dichalcogenide heterostructures[7], electrons and holes residing in different monolayers can bind into spatially indirect excitons[1,3,8-11] with a strong potential for optoelectronics[11,12], valleytronics[1,3,13], Bose condensation[14], superfluidity[14,15], and moiré-induced nanodot lattices[16]. Yet these ideas require a microscopic understanding of the formation, dissociation, and thermalization dynamics of correlations including ultrafast phase transitions. Here we introduce a direct ultrafast access to Coulomb correlations between monolayers; phase-locked mid-infrared pulses allow us to measure the binding energy of interlayer excitons in $WSe_2/WS_2$ hetero-bilayers by revealing a novel 1$s$-2$p$ resonance, explained by a fully quantum mechanical model. Furthermore, we trace, with subcycle time resolution, the transformation of an exciton gas photogenerated in the $WSe_2$ layer directly into interlayer excitons. Depending on the stacking angle, intra- and interlayer species coexist on picosecond scales and the 1$s$-2$p$ resonance becomes renormalized. Our work provides a direct measurement of the binding energy of interlayer excitons and opens the possibility to trace and control correlations in novel artificial materials.**

In monolayers of transition metal dichalcogenides (TMDs), the confinement of electronic motion into two dimensions and the suppression of dielectric screening facilitate unusually strong Coulomb interaction[17-20]. This gives rise to excitons with giant binding energies of several hundred meV[17], small Bohr radii[18] and ultrashort radiative lifetimes[19]. When two monolayers are contacted with type-II band alignment, the conduction band minimum and the valence band maximum are located in two different layers[7]. Owing to their proximity, electron-hole ($e$–$h$) pairs in adjacent monolayers are still subject to strong mutual Coulomb attraction. Interband photoluminescence combined with theory has, indeed, provided evidence of interlayer excitons[8,21-23]. Because the composite electron and hole wavefunctions overlap only weakly in space, these excitons are long lived – a key asset for future applications[1,3,14-16,24]. Yet, the weak coupling to light renders these quasiparticles inaccessible to interband absorption spectroscopy. Hence the binding energies of interlayer excitons, which depend sensitively on the delocalization of the electronic wavefunctions over the heterostructure[23], have not been measured. Signatures of the ultrafast interlayer charge transfer have been studied by interband spectroscopy[8,21]. These techniques, however, cannot measure Coulomb correlations or the formation of interlayer excitons, on the intrinsic ultrashort timescales.

Meanwhile, phase-locked electromagnetic pulses in the terahertz (THz) and mid-infrared (MIR) range have directly accessed ultrafast low-energy excitations, such as plasmons, phonons, or correlation-induced energy gaps[25-28]. By probing the internal $1s$-$2p$ resonance, pre-existing exciton populations have been studied, even if interband selection rules render them optically dark[19,20,26,27]. Whereas THz and MIR pulses have unveiled excitonic correlations in quantum wells[26], perovskites[27], and monolayer TMDs[19,20], excitons in TMD hetero-bilayers have not been resolved. As the library of two-dimensional van der Waals materials is rapidly growing[24], and exciting new phases may be tailored artificially[1-5], a precise quantitative understanding of Coulomb correlations between monolayers is vital.

Here we exploit near-infrared pump-MIR probe spectroscopy to measure the binding energy of spatially indirect excitons and the ultrafast transition of intralayer excitons into an interlayer insulating phase in $WSe_2$/$WS_2$ bilayers (Fig. 1a), depending on the stacking angle. The samples are manufactured by mechanical exfoliation of individual monolayers and subsequent deterministic transfer onto a diamond substrate (see Methods). In a first sample, monolayers of WSe$_2$ and WS$_2$ overlap in a large patch (Fig. 1b) of a diameter of ~80 μm with a stacking angle of θ = 5° ± 3° (see Methods). In the overlap area, the photoluminescence at the 1s A exciton lines of the bare monolayers is quenched (Fig. 1c), attesting to a good contact between the monolayers. We also prepared a reference monolayer of WSe$_2$ on a diamond substrate, covered with hexagonal boron nitride (hBN) (Supplementary Fig. S1a). This structure features a spatially uniform photoluminescence of the WSe$_2$ 1s A exciton (Supplementary Fig. S1b).

We first prove that MIR probing can unequivocally distinguish between spatially direct and indirect excitons. A 100-fs near-infrared (NIR) pump pulse, centred at a photon energy of 1.67 eV (Supplementary Fig. S2a) resonantly injects 1s A excitons in the WSe$_2$ monolayer (Fig. 1d). Since this energy lies well below the lowest interband resonance of WS$_2$ and hBN, we do not directly excite these materials. After a variable delay time $t_{pp}$, a phase-locked few-cycle MIR pulse (Fig. 1d, red waveform; centre frequency, 22 THz; spectral width, 19 THz, see also Supplementary Fig. S2b,c) is transmitted through the sample. The interaction with the photoinjected e–h pairs imparts a characteristic change on the amplitude and the phase of the MIR waveform, which we directly record by electro-optic sampling[19,20,25-28]. A Fourier transform and a Fresnel analysis allow us to extract the full dielectric response of the non-equilibrium system[25,26] (see Methods). The pump-induced changes of the real parts of the optical conductivity $\Delta\sigma_1$, and of the dielectric function $\Delta\varepsilon_1$, describe the absorptive and the inductive response, respectively.

Coulomb correlations leave characteristic fingerprints in the dielectric response. At $t_{pp}$ = 175 fs, the response in the reference sample is dominated by a maximum in $\Delta\sigma_1$ (Fig. 2a) and a corresponding zero crossing in $\Delta\varepsilon_1$ (Fig. 2b), at a photon energy of 144 ± 6 meV. This resonance marks the recently discovered 1s-2p transition of excitons within the WSe$_2$ monolayer[19]. The energy is slightly red-shifted with respect to the value of a bare WSe$_2$ monolayer[19], owing to dielectric screening by the hBN cover layer[20]. Most remarkably, an analogous experiment with the WSe$_2$/WS$_2$ heterostructure yields a dramatically different result (Fig. 2c,d). At a delay time of 5.1 ps after photoinjection of excitons in the WSe$_2$ monolayer, the strong resonance at an energy of 144 meV is absent. Instead, a maximum in $\Delta\sigma_1$ (Fig. 2c) and a dispersive feature in $\Delta\varepsilon_1$ (Fig. 2d) at an energy of 67 ± 6 meV dominate the response. This signature, which is spectrally far below the known internal 1*s*-2*p* resonances of intralayer excitons (see Methods) and far above the phonon frequencies of the TMD materials used here (see Supplementary Fig. S2d), is characteristic of interlayer excitons as shown next.

Owing to the type-I band alignment between WSe$_2$ and hBN (Fig. 2b, inset), excitons photoinjected in the reference sample remain within WSe$_2$[20]. The staggered band alignment in the WSe$_2$/WS$_2$ bilayer (Fig. 2d, inset), in contrast, should enable interlayer electron tunnelling[29]. This process competes with the radiative recombination of the intralayer excitons[19], quenching the photoluminescence of the WSe$_2$ monolayer[8] (Fig. 1c). The Coulomb attraction between electrons and holes remains strong even if they are spatially separated in neighbouring monolayers. The eigenenergies $E_b$ and wavefunctions $\varphi_\mathbf{q}$ of the bound states are obtained by solving the Wannier equation

$$\frac{\hbar^2 q^2}{2\mu}\varphi_\mathbf{q} - \sum_\mathbf{k} V_{\mathbf{k}-\mathbf{q}}\varphi_\mathbf{k} = E_b \varphi_\mathbf{q} \qquad (1)$$

Here μ is the reduced mass, and the effective 2D Coulomb potential $V_\mathbf{k}$ is derived by generalizing the Keldysh potential of the monolayer case to heterostructures and solving the Poisson equation self consistently. With literature values for the interlayer distance and the relevant dielectric constants, we compute a 1*s*-2*p* energy separation of 147 ± 3 meV and 69 ± 5 meV (Fig. 2e,f) for the reference sample and the WSe$_2$/WS$_2$ bilayer (see Methods), respectively, which coincide very well with the measured resonances of Fig. 2a-d. The 1*s* and 2*p* wavefunctions are qualitatively similar for intra- (Fig. 2e) and interlayer species (Fig. 2f), and the dipolar nature of the 2*p* wavefunctions warrants an in-plane orientation of the 1*s*-2*p* transition dipole for both cases. Because these resonances carry by far the largest oscillator strength of all intra-excitonic transitions they are expected to dominate the MIR spectra, as indeed observed experimentally. The excellent experiment-theory agreement allows us to quantify the corresponding binding energy of the interlayer excitons in the WSe$_2$/WS$_2$ bilayer under these conditions as 126 ± 7 meV (see Methods and Supplementary Information, section 3), substantially exceeding the thermal energy at room temperature.

Next, we utilize the hallmark MIR resonances to track the dynamics of Coulomb correlations. By repeating the NIR pump-MIR probe experiment for variable delay times $t_{pp}$, we map out the time evolution of $\Delta\sigma_1$ of the WSe$_2$/WS$_2$ heterostructure (Fig. 3a, see Supplementary Fig. S4e for $\Delta\varepsilon_1$). $\Delta\sigma_1$ sets on sharply at $t_{pp}$ = 0 ps. Only 100 fs later, a broad maximum at an energy of ~150 meV indicates the presence of intralayer excitons in WSe$_2$. Already at this time, a broad shoulder at ~67 meV attests to the formation of interlayer excitons. Within the subsequent 100 fs, the spectral weight shifts from the intra- to the interlayer resonance, as seen at $t_{pp}$ = 0.2 ps. Subsequently, the intralayer resonance disappears completely while the spectral weight of the interlayer exciton resonance continues to grow until $t_{pp}$ = 5.1 ps. This dynamics is faithfully reproduced by a microscopic model (Fig. 3c) based on the density matrix formalism discussed in more detail below (see also Methods). Our experimental data, thus, directly resolve the transformation of intralayer excitons into interlayer bound states in the time domain. For $t_{pp}$ > 5.1 ps, $\Delta\sigma_1$ decreases as the interlayer exciton population decays (Fig. 3a). Intriguingly, the exciton resonance is blue-shifted by ~20 meV, at $t_{pp}$ = 50 ps.

To test the influence of the stacking angle, we repeat the pump-probe experiment under identical conditions with a comparable WSe$_2$/WS$_2$ bilayer featuring θ = 27° ± 3° (see Supplementary Fig. S1c,d). The dynamics of $\Delta\sigma_1$ (Fig. 3b, see Supplementary Fig. S4f for $\Delta\varepsilon_1$) is similar to Fig. 3a, with two exceptions: (i) The spectral weight in the vicinity of the intralayer exciton resonance remains observable at $t_{pp}$ = 5.1 ps. (ii) The peculiar blue shift of the interlayer exciton resonance at $t_{pp}$ = 50 ps does not occur for θ = 27°.

For a quantitative discussion, we fit the MIR response with a phenomenological three-fluid model, which accounts for the two Lorentzian resonances of inter- and intralayer 1$s$-2$p$ transitions and includes a contribution of excited interlayer exciton states discussed below (see also Methods). The energies and dipole moments of the exciton resonances are adopted from the above theory whereas the inter- and intralayer exciton densities, $n_{inter}$ and $n_{intra}$, as well as the spectral weight of the excited interlayer exciton states are freely varied. The requirement to simultaneously reproduce the spectra of both $\Delta\sigma_1$ and $\Delta\varepsilon_1$ (Fig. 3a,b and Supplementary Fig. S4b,c,e,f) sets narrow boundaries for acceptable exciton densities (Fig. 4a,b). We observe that $n_{intra}$ (Fig. 4a) peaks right after the pump pulse, at $t_{pp}$ = 100 fs, for both heterostructures. The subsequent decay, however, is significantly slower in the strongly twisted bilayer (θ = 27°) than for θ = 5°, where a notable decay occurs within less than 0.2 ps. Furthermore, the maximum of $n_{intra}$ is larger for θ = 27°, even though the pump fluence is identical in both structures. This confirms that, for θ = 5°, tunnelling into the $WS_2$ layer occurs on the same timescale as photogeneration of intralayer excitons. Accordingly, $n_{inter}$ (Fig. 4b) rises quickly within the initial few hundred femtoseconds after optical excitation and levels off at $t_{pp}$ = 5.1 ps. The maximum observed for θ = 5° is approximately twice as large as in the more twisted bilayer, again attesting to a strongly enhanced tunnelling rate at θ = 5°. The MIR response of both hetero-bilayers can only be fitted satisfactorily if additional spectral weight below the observed frequency window is included. Such a response can be caused by unbound e–h pairs or excited-state absorption of bound states[30], both of which we phenomenologically account for in the form of a low-frequency Lorentz oscillator. From the temporal evolution of this component (Supplementary Fig. S5b,d,f) we estimate the density of unbound charge carrier to always remain at least a factor of 7 below the total density of e–h pairs, suggesting no strong intermediate plasma phase during the formation of interlayer excitons.

The observed dynamics is well described by a microscopic model illustrated in Fig. 4c. Because electrons and holes in interlayer excitons reside in different layers the lattice mismatch between $WSe_2$ and $WS_2$ leads to a momentum displacement between the intra- and the interlayer exciton dispersion (grey shaded parabolas). After photoinjection at a vanishing centre-of-mass momentum ($Q = 0$), 1s intralayer excitons spread out in reciprocal space via phonon scattering (Fig. 4c, orange arrow) while the entire ensemble is traced by its 1s-2p transition (Fig. 4c, blue arrows). The electrons can tunnel into the $WS_2$ layer under conservation of energy which is most efficient at the intersection points of the inter- and intralayer dispersion relations, where also the in-plane momentum $Q$ is conserved. Intralayer excitons may be transformed into interlayer unbound or bound states with a broad range of possible orbital quantum numbers. These quasiparticles subsequently relax towards the 1s interlayer state (Fig. 4c, green arrows), leading to the characteristic 1s-2p absorption (Fig. 4c, red arrows).

To model this scenario, we compute the temporal evolution of incoherent exciton densities for intra- and interlayer populations, including tunnelling, phonon scattering, radiative decay, non-radiative decay and intra-excitonic relaxation (see Methods). These simulations reproduce the spectral shape and the temporal evolution of $\Delta\sigma_1$ (Fig. 3c) as well as the dynamics of the intra- and interlayer exciton density, for $\theta = 5°$ and $27°$, at all $t_{pp} < 50$ ps (Fig. 4a,b, solid/dashed curves). Interestingly, direct tunnelling into the 1$s$ interlayer state is the most efficient quantum channel with a branching ratio of 54% (see Methods). Moreover, our model shows that the measurable difference of interlayer tunnelling rates of $\Gamma = (200 \text{ fs})^{-1}$ at $\theta = 5°$ and $\Gamma = (1.2 \text{ ps})^{-1}$ at $\theta = 27°$ is primarily caused by the momentum offset of inter- and intralayer exciton dispersions (see Methods).

Apart from the ultrafast formation of interlayer excitons, $\theta$ also strongly influences their long-term dynamics, where a blue shift of the 1$s$-2$p$ resonance occurs in the nearly aligned hetero-bilayer (Fig. 3a, for $t_{pp} = 50$ ps). We suggest that this phenomenon could be caused by lateral quantum confinement of excitons[16] (see also Supplementary Information, section 6). The slight lattice mismatch between WSe$_2$ and WS$_2$ entails a moiré pattern of alternating AA and AB stacking (Fig. 4d), which induces a periodic modulation of the potential energy experienced by interlayer excitons[16] (Fig. 4e). Owing to their small Bohr radii, the interlayer excitons may be captured in areas with AB stacking as they cool down. The additional quantum confinement could explain the blue shift of the 1$s$-2$p$ resonance for $t_{pp} = 50$ ps (Fig. 3a, see Methods).

This effect is even more pronounced in a third, precisely aligned hetero-bilayer with $\theta = 0° \pm 2°$ (see Supplementary Fig. S1e,f), where the moiré period reaches its largest value of 9 nm, the potential energy modulation is maximal, and interlayer excitons should be created with particularly small center-of-mass momenta owing to the minimum momentum mismatch between intra- and interlayer excitons. Indeed for this structure, the interlayer exciton resonance starts out at a relatively high photon energy of $83 \pm 6$ meV and does not shift any further for later delay times (see Supplementary Fig. S4a,d), suggesting that interlayer excitons may be efficiently captured in confined states already at early delay times $t_{pp}$. In contrast, the moiré period of 0.6 nm for $\theta = 27°$ (see Supplementary Fig. S1g), which is substantially smaller than the exciton Bohr radius (see Supplementary Fig. S1h), will not allow excitons to be efficiently captured, as indeed seen in Fig. 3b.

In conclusion, the direct study of interlayer Coulomb correlations allows us to measure the large binding energies of spatially indirect excitons. By probing all excitons, independently of their center-of-mass momenta, spin orientation or interband selection rules, we discover an unexpectedly efficient ultrafast transition of spatially direct into indirect excitons without a pronounced intermediate plasma phase of unbound $e$–$h$ pairs. The twist angle θ has a strong influence on the formation and cooling dynamics of interlayer excitons. While we extract a binding energy of $126 \pm 7$ meV for itinerant interlayer excitons, blue-shifted 1$s$-2$p$ resonances may be caused by quantum confinement within the moiré pattern. In the same spirit, observing Coulomb correlations and their ultrafast dynamics in the extensive library of van der Waals heterostructures opens a versatile playground for fundamental materials sciences and may become an invaluable help in searching for novel man-made equilibrium and hidden phases.

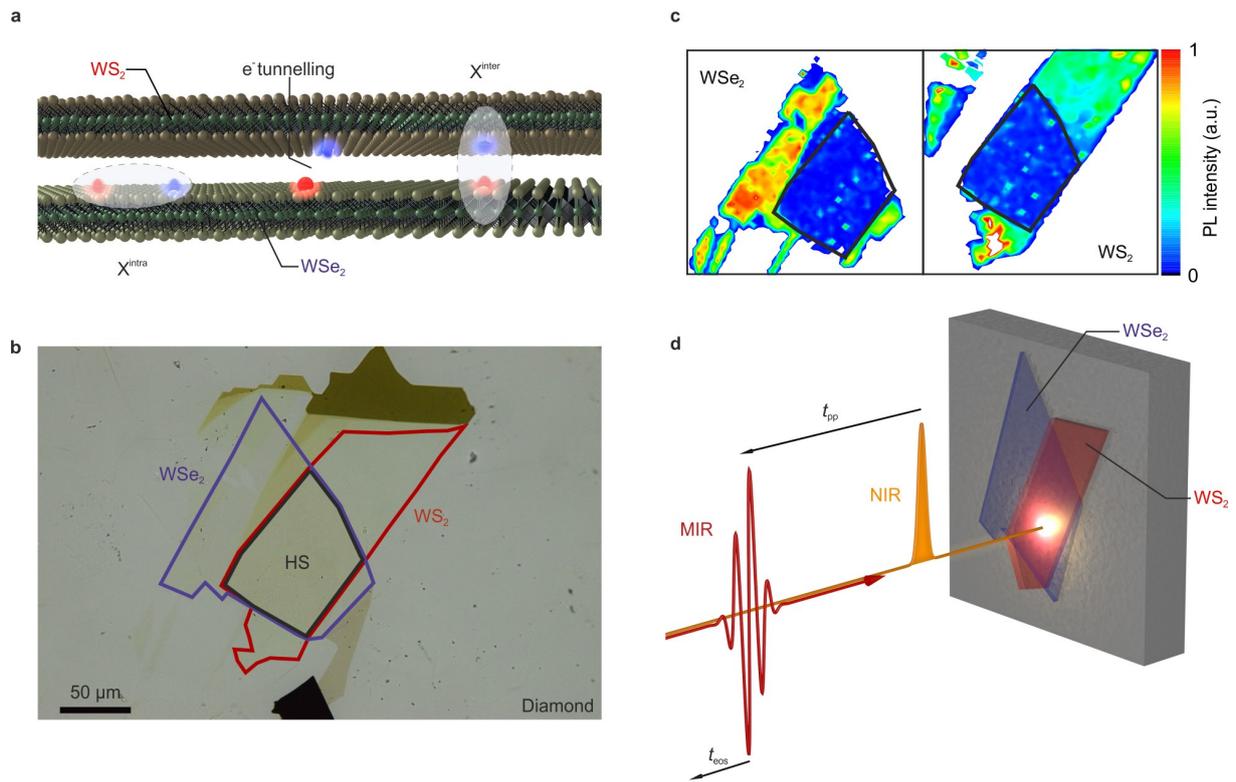

**Figure 1 | NIR pump-MIR probe spectroscopy of a WSe$_2$/WS$_2$ hetero-bilayer. a**, Artistic illustration of the electron tunnelling process and interlayer exciton formation out of intralayer excitons in a WSe$_2$/WS$_2$ heterostructure. **b**, Optical microscope image of the heterostructure. The purple frame indicates the WSe$_2$ monolayer, which is covered by a WS$_2$ monolayer (red frame). The heterostructure (HS) is formed in the overlap region (black frame). **c**, Micro-PL intensity map at the photon energy of the interband resonance of the 1*s* A exciton of WSe$_2$ (left panel) and WS$_2$ (right panel) when excited by a cw laser at a wavelength of 532 nm. All experiments are performed at room temperature. **d**, Time-resolved NIR pump-MIR probe spectroscopy of the heterostructure, which is van der Waals bonded on a diamond substrate. The 100-fs NIR pump pulse (orange) resonantly injects 1*s* A excitons in the WSe$_2$ monolayer. The ultrabroadband MIR probe pulse (red waveform) samples the dielectric response of the heterostructure.

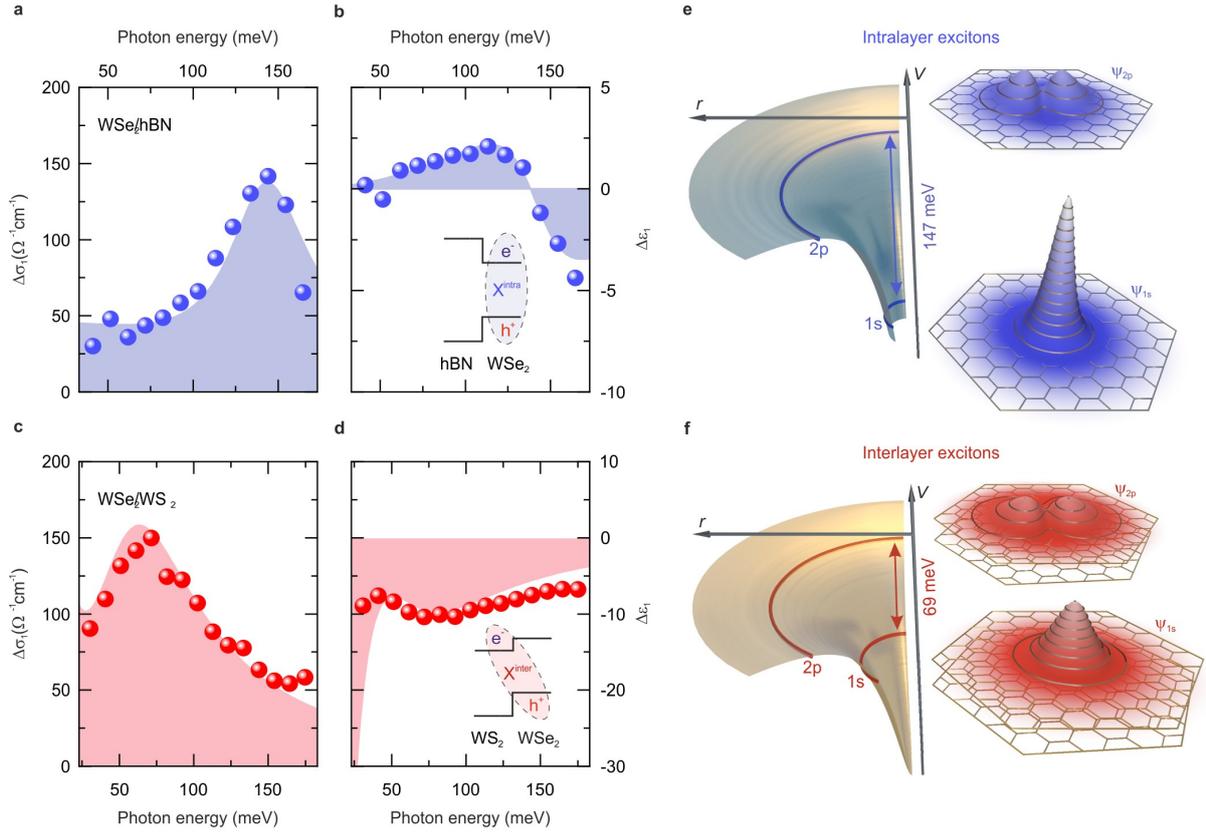

**Figure 2 | Dielectric response of intra- and interlayer excitons and excitonic wavefunctions. a-d**, Pump-induced changes of the real parts of the optical conductivity $\Delta\sigma_1$ (**a,c**), and the dielectric function $\Delta\varepsilon_1$ (**b,d**), as a function of the photon energy. The pump-induced dielectric response of the reference WSe$_2$/hBN structure is shown for $t_{pp}$ = 175 fs (**a,b**). For the $\theta$ = 5° WSe$_2$/WS$_2$ heterostructure (**c,d**), the response is depicted at $t_{pp}$ = 5.1 ps. The blue/red spheres denote the experimental data of the photoexcited heterostructures at a pump fluence of $\Phi$ = 27 μJ cm$^{-2}$. Shaded areas represent the three-fluid model fitting the experimental data as discussed in the main text. Inset in **b**: Schematic band alignment for the WSe$_2$/hBN heterostructure hosting intralayer excitons X$^{intra}$ in the WSe$_2$ monolayer. Inset in **d**: Band alignment for the WSe$_2$/WS$_2$ heterostructure supporting the formation of interlayer excitons X$^{inter}$. **e**, Intralayer Coulomb potential and wavefunctions of the 1$s$ and 2$p$ intralayer excitons in real space. **f**, Interlayer Coulomb potential and wavefunctions of the 1$s$ and 2$p$ interlayer excitons in real space.

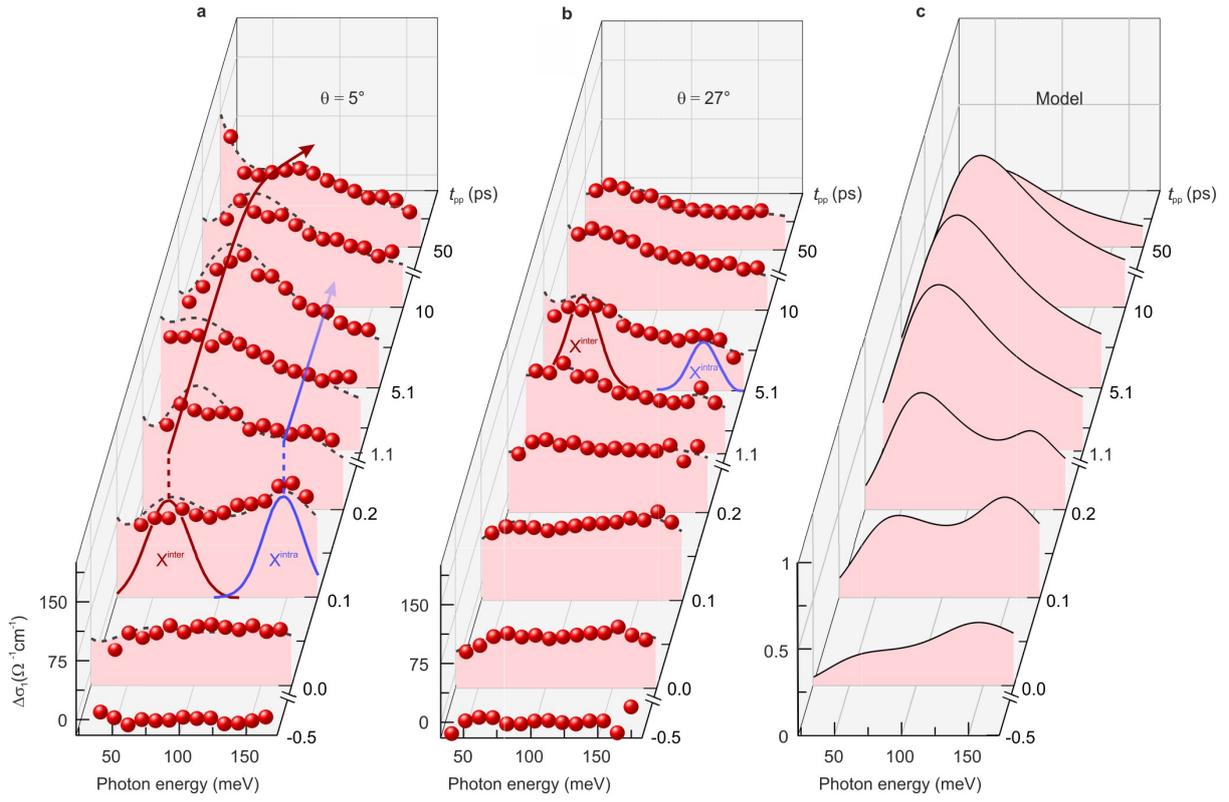

**Figure 3 | Temporal evolution of the dielectric response in experiment and theory. a,b,** Pump-induced change of the optical conductivity $\Delta\sigma_1$ of the photoexcited heterostructures, as a function of the photon energy at selected delay times $t_{pp}$ (pump fluence, $\Phi = 27$ µJ cm$^{-2}$), for the almost aligned ($\theta = 5°$) (**a**) and the misaligned ($\theta = 27°$) (**b**) heterostructure. Red spheres: experimental data. Dashed curves: phenomenological three-fluid model fitting the data as discussed in the main text. The red and blue Gaussian curves in **a** and **b** indicate the spectral positions of the inter- and intralayer exciton absorption. **c,** Normalized pump-induced change of the optical conductivity $\Delta\sigma_1$ modelled by the microscopic theory discussed in the main text, for $\theta = 5°$ and the delay times interrogated in the experiment.

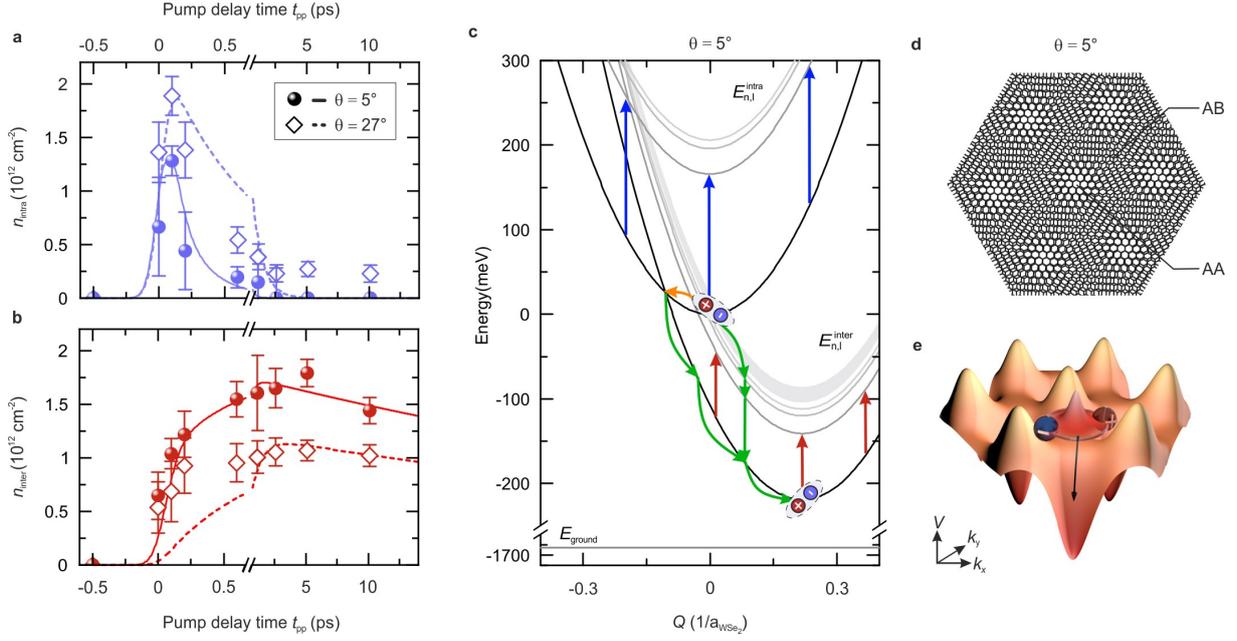

**Figure 4 | Ultrafast evolution of intra- and interlayer exciton densities and microscopic model. a,b**, Intralayer (**a**) and interlayer exciton density (**b**) extracted by the fitting procedure of the data in Fig. 3a,b and Supplementary Fig. S4b,c,e,f as a function of the pump-probe delay time $t_{pp}$ for the WSe$_2$/WS$_2$ heterostructures with θ = 5° (spheres) and θ = 27° (diamonds). The solid and dashed curves in **a** and **b** represent the result of the microscopic model discussed in the text, for θ = 5° and 27°, respectively. The error bars represent the 95% confidence interval of the fitting procedure. **c**, Dispersion of intra- and interlayer excitons in the two-particle picture, for θ = 5° and a number of orbital quantum states. Black parabolas: 1s excitons; grey parabolas: excitons with principal quantum numbers n > 1. The blue (red) arrows represent the 1s-2p transition of intralayer (interlayer) excitons. Phonon scattering can spread the intralayer exciton population in momentum space (orange arrow). The cascades of green arrows indicate possible decay channels for intralayer excitons to form interlayer states. The origin of the energy scale is shifted to the 1s intralayer exciton state with vanishing momentum Q. **d**, Superposition of the two monolayer lattices for a twist angle of θ = 5° forming a moiré pattern in real space. Bright regions correspond to AA stacking, whereas darker regions correspond to AB stacking. **e**, The real-space moiré pattern induces a local variation of the potential energy, whose period and amplitude strongly depend on the stacking angle. Small twist angles facilitate the localization of 1s interlayer excitons.

**Acknowledgements.** The authors thank Martin Furthmeier for technical assistance. This work was supported by the Deutsche Forschungsgemeinschaft (DFG) through research training group GRK 1570, collaborative research center SFB 1277 (Project A05), and project grant KO3612/3-1. The Chalmers group acknowledges funding from the European Union's Horizon 2020 research and innovation program under grant agreement No. 696656 (Graphene Flagship) and the Swedish Research Council (VR).

**Author Contributions** R.H. and E.M. supervised the study. P.M., F.M., P.S., A.G., and R.H. carried out the experiment. P.S., P.M., F.M., A.G., K.L., P.N., J.H., C.S., J.M.L., and T.K. prepared and pre-characterized the large-area heterostructures. S.O., S.B., and E.M. carried out the theoretical modelling. All authors analysed the data, discussed the results, and contributed to the writing of the manuscript.

**Author Information** Reprints and permissions information is available at www.nature.com/reprints. The authors declare no competing financial interests. Correspondence and requests for materials and data should be addressed to R.H. (rupert.huber@ur.de) or E.M. (ermin.malic@chalmers.se).

## Methods

**Sample fabrication and characterization.** The TMD monolayers are fabricated via mechanical exfoliation[31]. Starting from the bulk single crystal, a thin flake of the material is exfoliated with Nitto-tape and placed on a viscoelastic poly-dimethyl-siloxane (PDMS) substrate. By gently lifting the tape, a single monolayer of the TMD may remain on the substrate. The monolayer is pre-characterized under an optical microscope before being stamped onto a diamond substrate using a micro-positioning stage. This procedure can be repeated to stack multiple monolayers on top of each other while accurately aligning their position and twist angle θ with the translation stage. To confirm the structural integrity and contact of the heterostructure sample we record room-temperature micro-photoluminescence intensity maps of the individual monolayers after the transfer onto the heterostructure, excited by a cw laser at a wavelength of 532 nm. The angle θ is identified by tracing the characteristic cleavage edges along the armchair and zigzag direction of the monolayers spanning a 120° angle. Additionally, θ was confirmed for each heterostructure by polarization-resolved second harmonic generation[23,32]. Here, a linearly-polarized Ti:sapphire laser is used to excite the monolayers, and the polarization component of the second harmonic, parallel to the polarization of the fundamental, is analysed. This gives direct access to the armchair directions of the monolayers and therefore the twist angle of the hetero-bilayer.

**Ultrafast NIR pump-MIR probe spectroscopy.** Figure 1d depicts the experimental setup schematically. A 100-fs pump pulse centred at an energy of 1.67 eV (Supplementary Fig. S2a) from a home-built Ti:sapphire laser amplifier resonantly injects $1s$ A excitons in the $WSe_2$ monolayer at a repetition rate of 400 kHz. Another part of the laser output generates phase-locked MIR probe pulses by optical rectification in a 10 μm thick GaSe crystal. This probe pulse propagates through the photoexcited sample after a variable delay time $t_{pp}$. The MIR waveform and any changes induced by the sample are fully resolved by electro-optic sampling in a second GaSe crystal, as a function of the electro-optic sampling time $t_{eos}$ (Supplementary Fig. S2b). The MIR probe pulse is centred at a frequency of 22 THz with a full width at half maximum (FWHM) of 19 THz (Supplementary Fig. S2c. black curve) and a spectral phase which is flat between 15 and 45 THz (Supplementary Fig. S2c, blue curve). Using serial lock-in detection, we record the electric field of the MIR probe pulse propagating through the unexcited and the photoexcited sample. This allows us to determine the pump induced change $\Delta E(t_{eos})$ of the MIR electric field as function of the detection time $t_{eos}$. Applying a transfer matrix formalism, we have access to the full complex-valued dielectric response function, characterized by $\Delta\sigma_1$ and $\Delta\varepsilon_1$. This technique is sensitive to bound and unbound $e$–$h$ pair populations, irrespective of interband selection rules, and may thus interrogate optically dark and bright excitons[33].

**Solution of the Wannier equation and effective 2D Coulomb potential.** The excitonic binding energies and wavefunctions are obtained by finding the eigenenergies $E_b$, and eigenfunctions $\varphi_\mathbf{q}$ of the Wannier equation (see equation (1) in the main text). Here $\mathbf{q} = (q, \phi_q)$ is the electron-hole relative momentum in polar coordinates. The eigenfunctions can be expressed as $\varphi_\mathbf{q} = \varphi_q e^{im\phi_q}$, where m is the azimuthal quantum number. The effective Coulomb potential $V_\mathbf{q}(z)$ is obtained by solving the Poisson equation for a point charge placed at $z = 0$ in the dielectric environment

$$\varepsilon(z) = \begin{cases} \varepsilon_0 \varepsilon_{air} & z > \frac{1}{2}d_{WSe_2} + d_{gap} + d_{WS_2} \\ \varepsilon_0 \varepsilon_{WS_2} & \frac{1}{2}d_{WSe_2} + d_{gap} < z < \frac{1}{2}d_{WSe_2} + d_{gap} + d_{WS_2} \\ \varepsilon_0 \varepsilon_{gap} \quad \text{for} & \frac{1}{2}d_{WSe_2} < z < \frac{1}{2}d_{WSe_2} + d_{gap} \\ \varepsilon_0 \varepsilon_{WSe_2} & -\frac{1}{2}d_{WSe_2} < z < \frac{1}{2}d_{WSe_2} \\ \varepsilon_0 \varepsilon_{diamond} & z < -\frac{1}{2}d_{WSe_2} \end{cases}$$

We set the effective layer thicknesses $d_{WS_2} = d_{WSe_2} = 0.57$ nm and $d_{gap} = 0.1 \pm 0.03$ nm. For the relative permittivities we used $\varepsilon^\perp_{WSe_2} = 7.5$, $\varepsilon^\perp_{WS_2} = 6.3$, $\varepsilon^\parallel_{WSe_2} = 13.36$, $\varepsilon^\parallel_{WS_2} = 11.75$, $\varepsilon_{diamond} = 5$, $\varepsilon_{air} = 1$, and $\varepsilon_{gap} = 1$ (ref. 34, 35), where we account for the strong anisotropy of the dielectric tensor regarding the in- and out-of-plane direction of the adjacent TMD monolayers (see Supplementary Information section 3 for details). The resulting potentials $V_\mathbf{q}(z)$ for a test charge at $z = 0$ and $z = \frac{1}{2}d_{WSe_2} + d_{gap} + \frac{1}{2}d_{WS_2}$ were used as the intra- and interlayer potentials respectively. These, as well as the resulting 1$s$ and 2$p$ binding energies and wavefunctions, can be seen in Fig. 2e,f. While the interlayer exciton 1$s$-2$p$ transition energy clearly differs from the intralayer resonance, the latter is very similar for the hBN (147 ± 3 meV (ref. 20)) and the $WS_2$ (142 ± 9 meV) cover layer (see Supplementary Information section 3).

From here, we can calculate the oscillator strengths of the intra- and interlayer excitons as

$$f_{1s,2p} = \frac{2\mu}{\hbar^2} E_{1s,2p} |\langle \varphi^{1s} | \boldsymbol{r} \cdot \boldsymbol{e}_\parallel | \varphi^{2p} \rangle|^2$$

where $E_{1s,2p}$ is the 1s-2p transition energy and $\boldsymbol{e}_\parallel$ is the polarization of the MIR probe pulse. The oscillator strengths are 0.27 for intralayer excitons and 0.44 for interlayer excitons.

**Three-fluid model.** To analyse the measured MIR spectra more quantitatively, we apply a phenomenological three-fluid model. This approach describes the pump-induced change of the dielectric response function $\Delta\varepsilon(\omega) = \Delta\varepsilon_1(\omega) + \Delta\sigma_1(\omega)/(\varepsilon_0\omega)$ as follows:

$$\Delta\varepsilon(\omega) = \frac{n_{inter} e^2}{d\varepsilon_0 \mu_{inter}} \frac{f_{1s,2p}^{inter}}{\frac{E_{inter}^2}{\hbar^2} - \omega^2 - i\omega\Delta_{inter}} + \frac{n_{intra} e^2}{d\varepsilon_0 \mu_{intra}} \frac{f_{1s,2p}^{intra}}{\frac{E_{intra}^2}{\hbar^2} - \omega^2 - i\omega\Delta_{intra}} + \frac{e^2}{d\varepsilon_0 \mu_{inter}} \frac{A}{\frac{E_{higher}^2}{\hbar^2} - \omega^2 - i\omega\Gamma_{higher}} \quad (2)$$

where e is the electron charge, $\varepsilon_0$ denotes the vacuum permittivity, and $f_{1s,2p}^{inter}$ and $f_{1s,2p}^{intra}$ are the oscillator strengths of the 1s-2p transition in inter- and intralayer excitons calculated above. The first and the second term are Lorentzian resonances, representing the 1s-2p line of the inter- and intralayer excitons respectively. They include the interlayer/intralayer exciton density $n_{inter}/n_{intra}$, their resonance energy $E_{inter}/E_{intra}$ and their linewidth $\Delta_{inter}/\Delta_{intra}$. The reduced mass of the interlayer excitons $\mu_{inter} = m_e^{WS_2} m_h^{WSe_2}/(m_e^{WS_2} + m_h^{WSe_2}) = 0.15\, m_0$ is determined with the effective electron mass in the conduction band of $WS_2$, $m_e^{WS_2} = 0.27\, m_0$, and the effective hole mass in the valence band of $WSe_2$, $m_h^{WSe_2} = 0.36\, m_0$ (ref. 36). Here, $m_0$ is the free electron mass. The effective mass of the intralayer excitons $\mu_{intra} = 0.16\, m_0$ is determined with the electron and hole effective masses within the $WSe_2$ monolayer only[36]. The thickness of the sample (treated as a thin slab in this model) is set to $d = 0.7$ nm (ref. 19). The third term in equation (2) phenomenologically models all transitions between higher excited e–h pair states with an effective Lorentzian featuring a resonance energy of $E_{higher} = 4$ meV, and an effective scattering rate $\Gamma_{higher}$. $A = n_{higher} f_{higher}$ may be interpreted as an average density $n_{higher}$ of excited-state bound and unbound e–h pairs weighed by a mean oscillator strength $f_{\text{higher}}$. As shown in ref. 30, excited-state transitions in excitons and unbound e–h pairs are indeed expected to occur below the photon energy of our MIR probe pulses. To restrict the number of adjustable parameters, we include all information known *a priori*: The 1s-2p transition energies of the intra- and interlayer excitons are kept at the values found in Fig. 2 (with a maximum allowable variation of 15%) conforming to theory. Likewise, the inter- and intraexcitonic linewidths are kept between 40 and 120 meV (ref. 19, 20) and the excited state scattering time is set below 200 fs. Strict limits on the possible values of $n_{\text{inter}}$, $n_{\text{intra}}$, and $A$ are set by the fact that both independently retrieved spectra $\Delta\sigma_1$ and $\Delta\varepsilon_1$ have to be reproduced simultaneously. Without any further restrictions, the numerical adaption of the measured spectra yields an overall good fit quality (Supplementary Fig. S4, black dashed curves). The temporal evolution of the fitting parameters is given in Fig. 4a,b and Supplementary Fig. S5.

**Microscopic tunnelling model.** The model presented here is an extension of the one detailed in ref. 37, in which equations of motion for exciton densities are derived microscopically using density matrix theory[38] and the Heisenberg equation of motion in a second quantization formalism. It contains three main quantities: the intralayer microscopic polarization $P$, the intralayer exciton density $n_Q^{\text{intra}}$ and the interlayer exciton density $n_Q^{\text{inter}}$. Here $\boldsymbol{Q}$ is the electron-hole center-of-mass momentum. These quantities only describe the 1s K − K states, where the Coulomb-bound electron and hole are both located at the K valley in the Brillouin zone. The higher excitonic states (2s, 3s, 4s, etc.) are treated as a single quantity, $n^{\text{higher}}$. The exciton dynamics is described by the equations

$$\partial_t P = \Omega(t) - \left(\gamma_{\text{rad}}^{\text{intra}} + \gamma_P^{\text{intra}}\right) P \tag{3}$$

$$\partial_t n_Q^{\text{intra}} = 2\gamma_P^{\text{intra}} |P|^2 - 2\gamma_{\text{rad}}^{\text{intra}} \delta_{Q0} n_Q^{\text{intra}} + \partial_t n_Q^{\text{intra}}\big|_P + \partial_t n_Q^{\text{intra}}\big|_T \tag{4}$$

$$\partial_t n_Q^{\text{inter}} = -\left(2\gamma_{\text{rad}}^{\text{inter}} \delta_{Q0} + \gamma_{\text{nonrad}}^{\text{inter}}\right) n_Q^{\text{inter}} + \partial_t n_Q^{\text{inter}}\big|_P + \partial_t n_Q^{\text{inter}}\big|_T + \partial_t n_Q^{\text{inter}}\big|_{\text{higher}} \tag{5}$$

$$\partial_t n^{\text{higher}} = \partial_t n^{\text{higher}}\big|_T + \partial_t n^{\text{higher}}\big|_{\text{higher}} \tag{6}$$

In equation (3), $P$ is treated in its rotating frame, while $\Omega(t)$ is the Rabi frequency and describes the exciting optical pulse, which is modelled as a Gaussian with a FWHM of 100 fs. $\gamma_{rad}^{intra} = 2.59$ ps$^{-1}$ and $\gamma_P^{intra} = 26.6$ ps$^{-1}$ are the radiative and phonon-induced decay rates, respectively[37]. The first describes the radiative decay of excitons within the light cone. The latter leads to the formation of intralayer excitons as seen in the first term of equation (4). The second term in equation (4) describes the radiative decay of excitons within the light cone. The phonon scattering term $\partial_t n_Q^{intra}|_P = \sum_{Q'} \left( \Gamma_{P,Q'Q}^{intra} n_{Q'}^{intra} - \Gamma_{P,QQ'}^{intra} n_Q^{intra} \right)$ couples intralayer exciton states with different momenta. The two terms in the sum describe in- and out-scattering respectively, mediated by the intralayer phonon scattering rate $\Gamma_{P,QQ'}^{intra}$ which thermalizes the exciton distribution. Since $\Gamma_{P,QQ'}^{intra}$ is a matrix there is no single way to quantify it as a number. It can however be instructive to examine the angle-averaged out-scattering rate $\sum_{Q'} \Gamma_{P,QQ'}^{intra}$, which lies between 40 and 140 ps$^{-1}$. The last term in equation (4) is the tunnelling term $\partial_t n_Q^{intra}|_T = \sum_{Q'} \Gamma_{T,QQ'}^{intra-inter} \left( n_{Q'}^{inter} - (1+\alpha) n_Q^{intra} \right)$, coupling inter- and intralayer states with different momenta. The angle-averaged out-tunnelling rate $\sum_{Q'} \Gamma_{T,QQ'}^{intra-inte}$ lies between 1.4 and 6.8 ps$^{-1}$. Finally, the factor

$$\alpha = \frac{\sum_{Q',\mu \neq 1s} \Gamma_{T,0Q'}^{1s-\mu,intra-inter}}{\sum_{Q'} \Gamma_{T,0Q'}^{1s-1s,intra-int}} = 0.84$$

represents the excitons that tunnel into higher excitonic states. The form of $\alpha$ assumes that the appearing ratio does not change significantly with $Q$.

Equation (5) describes the dynamics of interlayer excitons. The two decay rates in the first term are given by $\gamma_{rad}^{inter} = 1.02 \times 10^{-6}$ ps$^{-1}$ (ref. 37) and $\gamma_{nonrad}^{inter} = 0.03$ ps$^{-1}$, where the latter is of phenomenological character. The phonon scattering term $\partial_t n_Q^{inter}|_P$ has the same form as for the intralayer exciton, with $\sum_{Q'} \Gamma_{P,QQ'}^{inter}$ lying between 6 and 40 ps$^{-1}$. Tunnelling is described by $\partial_t n_Q^{inter}|_T = \sum_{Q'} \Gamma_{T,QQ'}^{intra-inte} \left( n_{Q'}^{intra} - n_Q^{inter} \right)$ where $\Gamma_{T,QQ'}^{inter-intr} = \Gamma_{T,Q'Q}^{intra-int}$ which makes the 1s tunnelling symmetric.

The final term $\partial_t n_Q^{inter}|_{higher} = \frac{1}{\tau} \left( \beta n^{higher} \delta_{Q0} - (1-\beta) n_Q^{inter} \right)$ describes the interaction with the higher excitonic states. It relies on the assumption that the higher states scatter down to the 1s light cone

with zero momentum on a timescale $\frac{\beta}{\tau}$, where $\tau = 250$ fs. It also includes back scattering into the higher states on a timescale $\frac{1-\beta}{\tau}$. Here we choose $\beta = \frac{e^{-E_{\text{inter}}^{1s,K-K}/(k_BT)} + e^{-E_{\text{inter}}^{1s,K-K'}/(k_BT)} + e^{-E_{\text{inter}}^{1s,K-\Lambda}/(k_BT)}}{Z} = 0.57$ where $T$ is the temperature, $k_B$ is the Boltzmann constant and $Z$ is the canonical partition function. By doing this, the system is ensured to enter a Boltzmann distribution at equilibrium. Here, the $K - K'$ and $K - \Lambda$ states have been taken into account as well. This is motivated by our calculations of the binding energies revealing that the 1s-2p transition energy differs by less than 1 meV from that of the $K - K$ state.

Finally, equation (6) describes the dynamics of higher excitonic states. The first term $\partial_t n^{\text{higher}}|_T = \alpha \sum_{QQ'} \Gamma_{T,QQ'}^{\text{inter}-\text{intra}} n_{Q'}^{\text{intra}}$ describes tunnelling from the intralayer into interlayer excitons. The second term $\partial_t n^{\text{higher}}|_{\text{higher}} = \frac{1}{\tau}\left((1-\beta)\sum_{Q'} n_{Q'}^{\text{inter}} - \beta n^{\text{higher}}\right)$ handles the interaction with the 1s interlayer state.

To reproduce the densities shown in Fig. 4a,b we integrate over all momenta. Hence we find $n_{\text{inter}} = 0.90 \sum_Q n_Q^{\text{inter}}$, where the factor 0.90 compensates for the bleaching from the 2p states assuming a Boltzmann distribution. The corresponding factor for the intralayer excitons is considerably closer to 1.

In this model, only the tunnelling rate is affected by twisting[37]. In case of 27° twisting the energy-allowed momentum transfers for the $K - K$ transition lie close to those for the $K - K'$ and $K - \Lambda$ transitions. As a result the $\sum_{Q'} \Gamma_{T,QQ'}^{\text{inter}-\text{intr}}$ rate for 27° ends up between 0.2 and 0.9 ps$^{-1}$.

**Estimate of moiré confinement.** Here we introduce a perturbative treatment to approximate the influence of the periodic modulation of the potential landscape induced by the moiré pattern. Therefore, we solve the Schrödinger equation in polar coordinates $r$ and $\varphi$ assuming a circular finite potential well, meaning that

$$V(r) = \begin{cases} 0, & r \leq r_0 \\ V_0, & r > r_0 \end{cases}$$

For $V(r)$ we have chosen $V_0 = 0.1$ eV and $r_0 = 0.33\, a_m$, where $a_m = 3.4$ nm is the moiré period for a stacking angle of $\theta = 5°$ (ref. 39). Assuming the wavefunctions $\psi$ to be separable in $r$ and $\varphi$ we find the (unnormalized) bound-state solutions

$$\psi_{nm}(r,\varphi) = \begin{cases} J_m(k_+ r)e^{im\varphi}, & r \leq r_0 \\ \dfrac{J_m(k_+ r_0)}{K_m(k_- r_0)} K_m(k_- r)\, e^{im\varphi}, & r > r_0 \end{cases}$$

where $k_+ = \sqrt{\dfrac{2\mu}{\hbar^2} E_n}$ outside and $k_- = \sqrt{\dfrac{2\mu}{\hbar^2}(V_0 - E_n)}$ inside the potential well. The azimuthal quantum number is given by m, and $J_m$ and $K_m$ are the normal and modified Bessel functions of the first and second kinds. In the interval $0 \leq E_n \leq V_0$, n denotes the n'th solution to the equation:

$$E_n - \dfrac{V_0}{1 + \left(\dfrac{K_m(k_- r_0)\bigl(J_{m-1}(k_+ r_0) - J_{m+1}(k_+ r_0)\bigr)}{J_m(k_+ r_0)\bigl(K_{m-1}(k_- r_0) - K_{m+1}(k_- r_0)\bigr)}\right)} = 0$$

The corresponding binding energy for the 1s state is $E_{1s} = 20$ meV whereas the shallow troughs of the potential energy landscape do not confine the 2p state. This scenario is expected to result in a transition energy of $E_{2p} - E_{1s} = 20$ meV. Adding this solution to the unperturbed 1s-2p transition energy of the interlayer excitons of 67 meV describes the observed blue shift of the resonance energy by 20 meV.

**Data Availability Statement (DAS)**

The datasets generated during and/or analysed during the current study are available from the corresponding author on reasonable request.